%% file: phi1.tex
\documentclass[11pt,proc]{article}
\input article2.sty
\usepackage{graphicx}

\begin{document}
\input sanda.tex
\vspace*{4cm}
\rightline{DPNU-00-08}
\centerline{\Large On a clean determination of $\phi_2$}
\sk
\centerline{ A. I. SANDA}
\sk
\centerline{Department of Physics, Nagoya University}
\centerline{Nagoya, Japan}
\sk
\centerline{Abstract}
We point out that time dependent study of four decay modes
$B^0(t)\to \Dbar K_S$, $B^0(t)\to D K_S$
$\Bbar^0(t)\to \Dbar K_S$, and $\Bbar^0(t)\to D K_S$ allows us to measure
$\sin(\phi_1-\phi_2)$ - contrary to previously obtained result that
it measures $\sin(2\phi_1+\phi_3)$.
The time dependent study of $B_s^0(t)\to D^\pm_sK^\mp$, 
and the study of $B^-\to K^-(D,\Dbar)\to K^-f$ decay 
where $D\to f$ is a doubly Cabibbo suppressed decay 
yield information on $\phi_3$, as previusly expected. 
\vskip 2cm

\section{Introduction}
Large \cp~ violation in $B^0\to\psi K_S$ has been observed by Babar and Belle\cite{cp}.
Clean measurement of $\phi_1$ has been made. This is the first step toward a serious test of the standard model of \cp~violation. The statistical and systematic errors are large
but this will improve in due time. New CDF measurments are around
the corner. LHCB and BTeV is also comming up. Exciting program for \cp~violation study is on its way.
What is next? We would like to test the
unitarity relation $\phi_1+\phi_2+\phi_3=\pi$. How do we measure
$\phi_2$? For many years, \cp~ asymmetry in $B^0\to \pi\pi$ was thought to be the way to
measure $\phi_2$. The existence of penguin amplitudes which prohibit a theoretically
clean measurement was overcomed by the isospin analysis. But, recent 
indication\cite{ukai} that 
$Br(B^0\to\pi^0\pi^0)\sim 2\times 10^{-7}$, and the background associated with $\pi^0$ detection
 makes it very difficult to extract $\phi_2$ from the isospin analysis.
 
A Dalitz plot analysis of $B\to \rho\pi\to 3\pi$ has been 
proposed\cite{quinnsny}. While we
expect $B\to 3\pi$ to be dominated by $B\to \rho\pi$ which is a pure
\cp~ eigenstate, there is some oposite \cp~ admixture under the $\rho$ band in the Dalitz
plot. Further study of the Dalitz plot is necessary to see if this technique
can be used to extract $\phi_2$ without theoretical ambiguity.

\section{Determination of $\phi_2$ from $B^0,~\Bbar^0\to DK_S$ Decay}

Here we ask if there is a practical way to determine $\phi_2$ in a 
theoretically clean way. We point out that $\sin(\phi_1-\phi_2)$ can be determined without theoretical
ambiguity by studying two time dependent asymetries:
\ba
a(t)&=&\frac{\Gamma(\Bbar^0(t)\to DK_S)-\Gamma(B^0(t)\to D K_S)}
{\Gamma(\Bbar^0(t)\to DK_S)+\Gamma(B^0(t)\to D K_S)}\nn
\overline a(t)&=&\frac{\Gamma(\Bbar^0(t)\to \Dbar K_S)-\Gamma(B^0(t)\to \Dbar K_S)}
{\Gamma(\Bbar^0(t)\to \Dbar K_S)+\Gamma(B^0(t)\to \Dbar K_S)}.
\ea
Here, in addition to $B^0\to DK_S$, we can also use 
$B^0\to DK^{0*}\to DK_S\pi^0$, and $B^0\to D^{0*}K_S$. 

In the literature\cite{sanda}, it is stated that this asymmetry measures $\sin(2\phi_1+\phi_3)$
which of cause is identical to $-\sin(\phi_1-\phi_2)$ if we assume unitarity, $\phi_1+\phi_2+\phi_3=\pi$.
Another often made statement is that
$\arg(\V_{ub}^*\V_{ud}\V_{tb}\V_{td}^*)=\phi_1+\phi_3$.
The correct statement is that it is related to $\phi_2$ as given in \mref{2}.
One should remember not to use equalities which depend on a paticular
 phase convention,
for example,
$\arg(\V_{ub}^*\V_{ud})=\phi_3$ or $\arg(\V_{tb}\V^*_{td})=\phi_1$.
In the first round of theoretical analysis, we were mainly interested in identifing all 
possible \cp~asymmetries, and we did not
excercise sufficient care in deriving the expression for \cp~asymmetry
independent of the unitarity. 
The unitarity constaint, which we want to test, has crept into derivations.

To avoid using unitarity relation which we want to test, it is 
useful to express all physical observables in a rephasing invarint way,
and write the asymmetries in terms of $\phi_1,~\phi_2,~\phi_3$ 
using the definitions:
\ba
\phi_1 & = & \pi-\arg\left(\frac{-{\bf V}^*_{tb}{\bf V}_{td}}
{-{\bf V}^*_{cb}{\bf V}_{cd}}\right),\\
\phi_2 & = & \arg\left(\frac{{\bf V}^*_{tb}{\bf V}_{td}}
{-{\bf V}^*_{ub}{\bf V}_{ud}}\right),\mlab{2}\\
\phi_3 & = & \arg\left(\frac{{\bf V}^*_{ub}{\bf V}_{ud}}
{-{\bf V}^*_{cb}{\bf V}_{cd}}\right).
\mlab{angle}
\ea
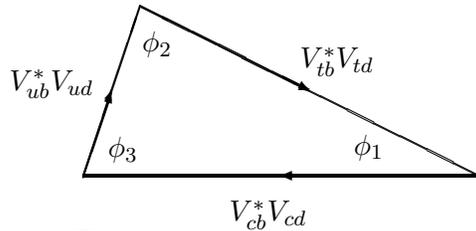
\begin{figure}[h]
\begin{picture}(400,160)(-30,210)
\put(80,300){\line(1,0){150}}
\put(80,300.5){\line(1,0){150}}
\put(230,300){\vector(-1,0){75}}
\put(230,300.5){\vector(-1,0){75}}
\put(150,285.5){\makebox(0,0){$V^*_{cb}V_{cd}$}}
\put(80,300){\line(1,3){21.5}}
\put(80,300.5){\line(1,3){21.5}}
\put(80,299.5){\line(1,3){21.5}}
\put(80,300){\vector(1,3){10.75}}
\put(80,300.5){\vector(1,3){10.75}}
\put(80,299.5){\vector(1,3){10.75}}
\put(68,335){\makebox(0,0){$V^*_{ub}V_{ud}$}}
\put(230,300){\line(-2,1){128}}
\put(230,300.5){\line(-2,1){128}}
\put(101.5,364.5){\vector(2,-1){64}}
\put(101.5,365){\vector(2,-1){64}}
\put(176,343.5){\makebox(0,0){$V^*_{tb}V_{td}$}}

\put(95,310){\makebox(0,0){$\phi_3$}}
\put(188,311){\makebox(0,0){$\phi_1$}}
\put(108,350){\makebox(0,0){$\phi_2$}}
\end{picture}
\vspace{-3cm}
\caption{\small The unitarity triangle in the complex plane.\label{unit}}
\end{figure}

The \cp~asymmetry for $\Bbar^0\to\psi K_S$ is theoretically clean at the level where
we neglect ${\cal O}(\sin^2\theta_c)$ terms. Here $\theta_c$ is the Cabibbo angle. 
This comes from the fact that, for $B^0\to\psi K_S$ decay,
 the tree graph, which is proportional
to $\V_{cb}\V_{cs}^*$, and the penguin graph, which is proportional to 
$\V_{tb}\V_{ts}^*$, have the same phase
if we neglect terms proportional to $\V_{ub}\V_{us}^*$. 
The latter neglected term is down by  ${\cal O}(\sin^2\theta_c)$
compared to the former.

The $2\times 2$ submatricies of the KM matrix are known to be approximately
unitary:
\ba 
{\bf V}^*_{ud}{\bf V}_{us}&\cong&-{\bf V}^*_{cd}{\bf V}_{cs}\nn
{\bf V}_{ud}{\bf V}^*_{cd}&\cong&-{\bf V}_{us}{\bf V}^*_{cs}\nn
{\bf V}^*_{cs}{\bf V}_{cb}&\cong&-{\bf V}^*_{ts} {\bf V}_{tb}.
\mlab{1}
\ea
These relations have been varified experimentally to within 6\%,
.1\%, 3\%, respectively, relative to terms that are kept\cite{PDG}. So, since 
we are making statments that are accurate to order $\sin^2\theta_c$,
we are free to use these relations 
 to test the unitarity relation:
\be 
{\bf V}_{td}{\bf V}^*_{ud} + {\bf V}_{ts}{\bf V}^*_{us} + {\bf V}_{tb} {\bf V}^*_{ub} =0 
\mlab{TRI6}  
\ee 

Now, let us discuss $B^0(t)\to \Dbar K_S$, $B^0(t)\to D K_S$,
$\Bbar^0(t)\to \Dbar K_S$, and $\Bbar^0(t)\to D K_S$ decays.
Defining $|K_S\ket$ state as
\be
|K_S\ket=p_K|K^0\ket+q_K|\Kbar^0\ket
\ee
and using \cp~symmetry of strong interaction, we obtain
\ba 
A( \overline B \to \overline D K_S) &=& 
e^{i\delta _-} {\bf V}_{cb}{\bf V}^*_{us}  {\cal A_-} q_K^*
\nn 
A( \overline B \to D K_S) &=& 
e^{i\delta _+} {\bf V}_{ub}{\bf V}^*_{cs}  {\cal A_+} q_K^*\nn 
A( B \to D K_S) &=& 
e^{i\delta _-} {\bf V}^*_{cb}{\bf V}_{us}  {\cal A_-} p_K^*\nn 
A( B \to \overline D K_S) &=&  
e^{i\delta _+} {\bf V}^*_{ub}{\bf V}_{cs}  {\cal A_+}p_K^*,
\ea 
where $\delta_\pm$ are stong phases, ${\cal A_\pm}$ are the magnitudes
of the matrix elements.

Note that $|A( \overline B \to DK_S)| = |A(B \to D K_S)| $ 
obviously does {\em not} hold as an identity. Setting 
$\Delta \Gamma_B =0$, for simplicity, we find 
\ba 
\Gamma(B(t)\to{DK_S}) &\propto& |A(DK_S)|^2 \bigg[ 
1 + |\overline \rho (DK_S)|^2 \nn
&& + (1 - |\overline \rho (DK_S)|^2) {\rm cos}\Delta M_B t 
-  2 {\rm Im}\left(\frac{q_B}{p_B} \overline \rho (DK_S) \right)
{\rm sin}\Delta M_Bt  \bigg] \qquad \nn
\Gamma(\Bbar(t)\to{\Dbar K_S}) &\propto& |\overline A(\Dbar K_S)|^2 \bigg[ 
1 + |\rho (\Dbar K_S)|^2\nn
&& + (1 - |\rho (\Dbar K_S)|^2) {\rm cos}\Delta M_B t - 
2 {\rm Im}\left( \frac{p_B}{q_B}\rho (\Dbar K_S) \right) {\rm sin}\Delta M_Bt  \bigg] \qquad\quad\nn
\Gamma(B(t)\to {\Dbar K_S}) &\propto& |A(\Dbar K_S)|^2 \bigg[ 
1 + |\overline \rho (\Dbar K_S)|^2\nn
&& + (1 - |\overline \rho (\Dbar K_S)|^2) {\rm cos}\Delta M_B t  
- 2 {\rm Im}\left(\frac{q_B}{p_B} \overline \rho (\Dbar K_S) \right)
{\rm sin}\Delta M_B t  \bigg] \qquad\nn
\Gamma(\Bbar(t)\to{DK_S}) &\propto& |\overline A(DK_S)|^2 \bigg[ 
1 + | \rho (DK_S)|^2\nn
&& + (1 - |\rho (DK_S)|^2) {\rm cos}\Delta M_B t - 
 2 {\rm Im}\left( \frac{p_B}{q_B}\rho (DK_S) \right) {\rm sin}\Delta M_B t  \bigg] \qquad\quad\nn
 \mlab{15}
 \ea
Using
\be
\frac{q_Bq_K^*}{p_Bp_K^*}=\frac{{\bf V}_{td}{\bf V}^*_{tb}}{{\bf V}^*_{td} {\bf V}_{tb}} 
\frac{{\bf V}_{cs}{\bf V}^*_{cd}}{{\bf V}^*_{cs} {\bf V}_{cd}} .
\ee
we obtain:
 \ba
\frac{q_B}{p_B}\overline \rho (\Dbar K_S) &=& e^{i(\delta _- - \delta _+)} 
\frac{{\bf V}_{cb}{\bf V}^*_{us}}{{\bf V}^*_{ub} {\bf V}_{cs}} 
\frac{{\bf V}_{td}{\bf V}^*_{tb}}{{\bf V}^*_{td} {\bf V}_{tb}} 
\frac{{\bf V}_{cs}{\bf V}^*_{cd}}{{\bf V}^*_{cs} {\bf V}_{cd}} \frac{ {\cal A}_-}{ {\cal A}_+}\nn 
\frac{q_B}{p_B}\overline \rho (D K_S) &=& e^{-i(\delta _- - \delta _+)} 
\frac{{\bf V}_{ub}{\bf V}^*_{cs}}{{\bf V}^*_{cb} {\bf V}_{us}} 
\frac{{\bf V}_{td}{\bf V}^*_{tb}}{{\bf V}^*_{td} {\bf V}_{tb}} 
\frac{{\bf V}_{cs}{\bf V}^*_{cd}}{{\bf V}^*_{cs} {\bf V}_{cd}}\frac{ {\cal A}_+}{ {\cal A}_-}\nn 
\ea

Note that this is rephasing invariant.
Using \mref{1}, we obtain:
\be
\frac{{\bf V}_{cb}{\bf V}^*_{us}}{{\bf V}^*_{ub} {\bf V}_{cs}} 
\frac{{\bf V}_{td}{\bf V}^*_{tb}}{{\bf V}^*_{td} {\bf V}_{tb}} 
\frac{{\bf V}_{cs}{\bf V}^*_{cd}}{{\bf V}^*_{cs} {\bf V}_{cd}} 
=-\frac{{\bf V}_{cb}{\bf V}^*_{cd}}{{\bf V}^*_{ub} {\bf V}_{ud}} 
\frac{{\bf V}_{td}{\bf V}^*_{tb}}{{\bf V}^*_{td} {\bf V}_{tb}} 
\ee
and using \mref{angle}:
\ba
\frac{q_B}{p_B}\overline \rho (\Dbar K_S) &=& -e^{i(\delta _- - \delta _+)} 
\frac{ {\cal A}_-}{ {\cal A_+}}\frac{1}{{\cal R}} e^{i(\phi_2-\phi_1)}\nn 
\frac{q_B}{p_B}\overline \rho (D K_S) &=& -e^{-i(\delta _- - \delta _+)} 
\frac{ {\cal A_+}}{ {\cal A}_-}{\cal R} e^{i(\phi_2-\phi_1)}\nn 
\ea
where 
\be
{\cal R}=\left|\frac{{\bf V}_{ub}{\bf V}^*_{ud}}{{\bf V}^*_{cb} {\bf V}_{cd}} 
\right|
\ee
Measurementing the coefficients of $\cos\Delta M_Bt$,
we can determine $\frac{ {\cal A_+}}{ {\cal A}_-}{\cal R}$.
The the coefficient of $\sin\Delta M_Bt$ gives $\sin(\phi_1-\phi_2)$.

\section{Determination of $\phi_3$ with $B^-\to K^-D$ decays}
It is useful to reanalize the other well known methods to determine $\phi_3$.
Measurements of $Br(B^-\to K^-(D^0,\Dbar^0)\to K^-f$, the
doubly Cabibbo suppressed decay $Br(D^0\to f)$, $Br(\Dbar^0\to f)$, 
$Br(B^-\to K^-D^0)$, and $Br(B^-\to K^-\Dbar^0)$ allow us to 
determine\cite{soni}
$\phi_3$.
Consider 
\be
B^-\to K^-(D^0,\Dbar^0)\to K^-f
\ee
Write
\ba
{\cal A}&=&A(B^-\to D^0K^-)A(D^0\to f)\nn
\overline{\cal A}&=&A(B^-\to \Dbar^0K^-)A(\Dbar^0\to f)
\ea
where these amplitudes represent strong interaction part and the KM 
factors are taken out.
\be
A(B^-\to K^-(D^0,\Dbar^0)\to f K^-)=[|{\cal A}|e^{i\delta}\V_{cb}\V_{us}^*\V_{us}\V_{cd}^*
+|\overline{\cal A}|e^{i\overline\delta}\V_{ub}\V_{cs}^*\V_{ud}^*\V_{cs}]
\mlab{tri}
\ee
Here $\delta$ and $\overline\delta$ are strong interaction phases for 
decays involving $D$ and $\Dbar$ intermediate particles, respectively.
It should be noted that by choosing the doubly Cabibbo suppressed decay
of $D$ meson, and that $B^-\to K^-\Dbar$ is color suppressed, 
the two terms on the right hand side have approximately
equal magnitude.

The relative phase of these amplitudes is given by
\be
\frac{A(B^-\to K^-D^0\to K^-f)}{A(B^-\to K^-\Dbar^0\to K^-f)}
=\frac{e^{i\delta}\V_{cb}\V_{us}^*\V_{cd}^*\V_{us}}{e^{i\overline\delta}\V_{ub}\V_{cs}^*\V_{ud}^*\V_{cs}}\frac{|{\cal A}|}{|\overline{\cal A}|}
\mlab{19}
\ee
Noting that 
\ba
A_1&=&\sqrt{Br(B^-\to K^-(D^0,\Dbar^0)\to K^-f)},\nn
A_2&=&\sqrt{Br(B^-\to D^0K^-)Br(D^0\to f)}\propto|{\cal A}\V_{cb}\V_{us}^*\V_{us}\V_{cd}^*|\nn
A_3&=&\sqrt{Br(B^-\to \Dbar^0K^-)Br(\Dbar^0\to f)}\propto|\overline{\cal A}\V_{ub}\V_{cs}^*\V_{ud}^*\V_{cs}|,
\ea
we can draw a triangle whose base is $A_1$
 and the two
sides are given by $A_2$ and 
$A_3$. \mref{19} gives the angle at the top of the triangle:
\be
\phi=\pi-\arg\left[\frac{e^{i\delta}\V_{cb}\V_{cd}^*}{e^{i\overline\delta}\V_{ub}\V_{ud}^*}\right]
=-\phi_3+\overline\delta-\delta
\ee
\begin{figure}[t]

\centerline{\psfig{figure=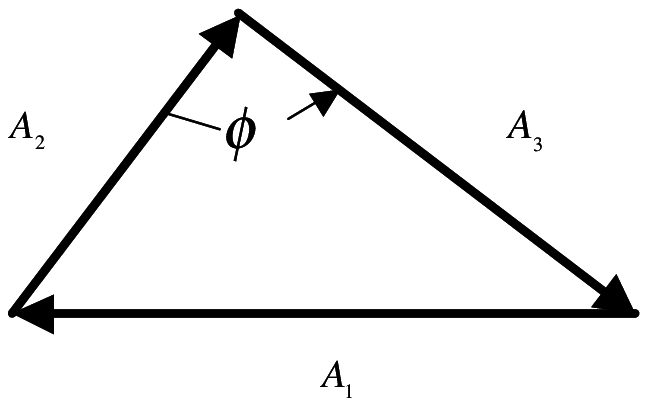,height=1.5in}}
\caption{
$A_1=|A(B^-\to K^-(D^0,\Dbar^0)\to K^-f)|$, $A_2=|A(B^-\to D^0K^-)A(D^0\to f)|$
and $A_3=|A(B^-\to \Dbar^0K^-)A(\Dbar^0\to f)|$ form a triangle. 
The angle at the top of the triangle $\phi=-\phi_3+\overline\delta-\delta$. The analysis of the charge conjugate modes give $\overline\phi=-\phi_3-\overline\delta+\delta$.
Thus $\phi_3$ can be determined.
decay.\label{bdecay}}
\end{figure}

The same analysis for $B^+$ decay leads to the determination
of $\phi_3+\overline\delta-\delta$ so both $\phi_3$ and the relative
strong interaction phase $\delta-\overline\delta$ can be determined.

\section{Determination of $\phi_3$ with $B_s^0\to D^\pm_sK^\mp$ decays} 
Both $B_s\to D^\pm_sK^\mp$ and $\Bbar_s^0\to D^\pm_sK^\mp$ exist
and there will be \cp~asymmetry which is expected to lead to information on 
$\phi_3$. We start be denoting relevant amplitudes:
\ba 
A( \overline B_s\to  D_s^- K^+) &=& 
e^{i\delta _-} {\bf V}_{ub}{\bf V}^*_{cs}  {\cal A_-} 
\nn 
A( \overline B_s\to D_s^+ K^-) &=& 
e^{i\delta _+} {\bf V}_{cb}{\bf V}^*_{us}  {\cal A_+} \nn 
A( B_s\to D_s^- K^+) &=& 
e^{i\delta _+} {\bf V}^*_{cb}{\bf V}_{us}  {\cal A_+} \nn 
A( B_s\to D_s^+ K^-) &=&  
e^{i\delta _-} {\bf V}^*_{ub}{\bf V}_{cs}  {\cal A_-},
\ea 
where $\delta_\pm$ are stong phases, ${\cal A_\pm}$ are the magnitudes
of the matrix elements.

\be
\frac{q_{B_s}}{p_{B_s}}\frac{A(\Bbar_s\to D_s^+K^-)}{A(B_s\to D_s^+K^-)}
=\frac{e^{i\delta_+}\V_{tb}^*\V_{ts}}{e^{i\delta _-}\V_{tb}\V_{ts}^*}
\frac{\V_{us}^*\V_{cb}}{\V_{cs}\V_{ub}^*}\frac{\cal A_+}{\cal A_-}
\mlab{12}
\ee
\ba
\frac{\V_{tb}^*\V_{ts}}{\V_{tb}\V_{ts}^*}
\frac{\V_{us}^*\V_{cb}}{\V_{cs}\V_{ub}^*}
&\cong&
\frac{\V_{cb}^*\V_{cs}}{\V_{cb}\V_{cs}^*}
\frac{\V_{us}^*\V_{cb}}{\V_{cs}\V_{ub}^*}\nn
&=&
\frac{\V_{cb}^*\V_{us}^*}{\V_{cs}^*\V_{ub}^*}
\frac{\V_{cs}\V_{ud}}{\V_{cs}\V_{ud}}\nn
&\cong&-\frac{\V_{cb}^*\V_{cd}}{\V_{cs}^*\V_{cs}}
\frac{\V_{ud}^*\V_{ud}}{\V_{ub}^*\V_{ud}}\nn
&\cong&\frac{1}{\cal R}e^{-i\phi_3}
\ea

\be
\frac{q_{B_s}}{p_{B_s}}\frac{A(\Bbar_s\to D_s^+K^-)}{A(B_s\to D_s^+K^-)}
=\frac{1}{{\cal R}}\frac{\cal A_+}{\cal A_-}e^{i(\delta_+-\delta_-)-i\phi_3}
\ee
Using \mref{1}, the right hand side of \mref{12} can be written as
Using the time dependence similar to \mref{15}, we can obtain both
the magnitude and the phase of $\frac{q_{B_s}}{p_{B_s}}\frac{A(\Bbar_s\to D_s^+K^-)}{A(B_s\to D_s^+K^-)}$. Similarly we can obtain  the magnitude and the phase
of 
\ba
\frac{q_{B_s}}{p_{B_s}}\frac{A(\Bbar_s\to D_s^-K^+)}{A(B_s\to D_s^-K^+)}
&=&\frac{e^{i\delta_-}\V_{tb}^*\V_{ts}}{e^{i\delta _+}\V_{tb}\V_{ts}^*}
\frac{\V_{ub}\V_{cs}^*}{\V_{cb}^*\V_{us}}\mlab{12}\frac{\cal A_-}{\cal A_+}\\
&=&{\cal R}\frac{\cal A_-}{\cal A_+}e^{i(\delta_--\delta_+)-i\phi_3}.
\ea
Thus $\phi_3$ can be determined.

\section{Summary}
The time has come to perform a serious test of the KM ansatz for \cp~violation. To see if the
unitarity triangle is actually a triangle is the next step. 
For this purpose we should be careful in deriving the expression for asymmetry 
in terms of three angles of the unitarity triangle. A useful rule is to always
make sure that the asymmetry is indeopendent of phase conventions.
We have reanalyzed processes known to yeild information on $\phi_3$.
Asymmetries for $B^0,~\Bbar^0\to DK_S$ decays are shown to yield information
on $\phi_2-\phi_1$, and not on $2\phi_1+\phi_3$ as previously stated. 
The other modes analyzed yield information on $\phi_3$
as expected.

\sk
{\bf Acknowledgments}
\sk
The author's research is supported by Grant-in Aid for Special Project Research
(Physics of CP violation). I thank T. Kurimoto,
Y. Okada, K. Ukai, and M. Yamauchi for useful discussions.
\sk
{\bf References}
\sk

\end{document}

%% file: sanda.tex
\def\bfix{~\newline\centerline{XXXXXXXXXXX corrected versionXXXXXXXXXXXXXX}\newline}
\def\efix{~\newline\centerline{XXXXXXXXXXXX corrected version ends hereXXXX}\newline}
\def\bfix{}
\def\efix{}
\def\bXX{}
\def\eXX{}
\def\ket{{\rangle}}
\def\bra{{\langle}}
\def\ie{{\it i.e.}}
\def\be{\begin{equation}}
\def\ee{\end{equation}}
\def\ba{\begin{eqnarray}}
\def\ea{\end{eqnarray}}
\def\bq{\begin{quotation}\noindent}
\def\eq{\end{quotation}}
\def\mref#1{Eq. (\ref{Eq:#1})}
\def\mreff#1{Fig. \ref{Eq:#1}}
\def\mreft#1{Table \ref{Eq:#1}}
\def\mlab#1{\label{Eq:#1}}
\def\mlabf#1{\label{Eq:#1}}
\def\mlabt#1{\label{Eq:#1}}
\def\half{\frac{1}{2}}
\def\to{\rightarrow}
\def\nn{\nonumber\\}
\def\sk{\vskip 1cm}
\def\skk{\vskip 3mm}
\def\mat#1#2#3{\langle{#1}\vert{#2}\vert{#3}\rangle}
\def\etal{{\it et al.}}
\def\etc{{\it etc.~}}

\def\ibid#1#2#3{{\it ibid. }{\bf #1} #2 {(#3)}}
\def\PR#1#2#3 {{\it Phys. Rev. }{\bf D#1} #2 {(#3)}}
\def\PRL#1#2#3 {{\it Phys. Rev. Lett. }{\bf #1} #2 {(#3)}}
\def\PL#1#2#3 {{\it Phys. Lett. }{\bf #1} #2 {(#3)}}
\def\AP#1#2#3 {{\it Ann, Phys. }{\bf #1} #2 {(#3)}}
\def\ZP#1#2#3 {{\it Z. Phys. }{\bf #1} #2 {(#3)}}
\def\NP#1#2#3 {{\it Nucl. Phys. }{\bf #1} #2 {(#3)}}
\def\MPL#1#2#3 {{\it Mod. Phys. Lett. }{\bf #1} #2 {(#3)}}
\def\NC#1#2#3 {{\it Nuov. Cim. }{\bf #1} #2 {(#3)}}
\def\PREP#1#2#3 {{\it Phys. Report }{\bf #1} #2 {(#3)}}
\def\PROG#1#2#3 {{\it Prog. Theor. Phys. }{\bf #1} #2 {(#3)}}
\def\SOV#1#2#3{{\it Sov. J. Nucl. Phys. }{\bf #1} #2 {(#3)}}
\def\JETP#1#2#3{{\it JETP }{\bf #1} #2 {(#3)}}
\def\RMP#1#2#3{{\it Rev. Mod. Phys. }{\bf #1} #2 {(#3)}}

\def\phiout{{\phi_a^{out}}}
\def\phiin{\phi_a^{in}}
\def\psiout{\psi_a^{out}}
\def\psiin{\psi_a^{in}}
\def\vin{\phi_{a\mu}^{in}}
\def\vout{\phi_{a\mu}^{out}}
\def\ofx{{(x)}}
\def\op{{\bf P}}
\def\oc{{\bf C}}
\def\ot{{\bf T}}
\def\cp{{\bf CP}}
\def\cpt{{\bf CPT}}
\def\vecr{{\vec r}}
\def\vecn{{\vec {\nabla}}}
\def\vecA{{\vec A}}
\def\psibar{\overline\psi}
\def\outin{{{out}\choose{in}}}
\def\inout{{{in}\choose{out}}}

\def\vr{\vec x}
\def\itt{{\it T}}
\def\b{{\bf b}}
\def\a{{\bf a}}
\def\d{{\bf d}}

\def\alphadot{{\dot\alpha}}
\def\betadot{{\dot\beta}}
\def\gammadot{{\dot\gamma}}

\def\N{\sqrt{{{E+m}\over{2m}}}}
\def\F#1{{{#1}\over{E+m}}}
\def\itemm{\hangindent\parindent\textindent}
\def\noi{\noindent}
\def\onehead#1{\vskip1pc\leftline{\bf #1}}
\def\twohead#1{\vskip1pc\leftline{\bf #1}}
\def\ts{\thinspace}

\def\sq2{{1\over{\sqrt{2}}}}
\def\omegaar{{\vec{\omega}}}
\def\kbar{\overline K}
\def\Pbar{\overline P}
\def\Abar{\overline A}
\def\dbar{\overline d}
\def\ubar{\overline u}
\def\sbar{\overline s}
\def\nubar{\overline \nu}
\def\cbar{\overline c}
\def\Dbar{\overline D}
\def\Bbar{\overline B}
\def\Kbar{\overline K}

\def\bbar{\overline B}
\def\gbar{\overline g}
\def\pbar{\overline P}
\def\qbar{\overline q}
\def\vecr{\vec r}

\def\dm{\Delta m}
\def\ss{(1+s^2)}
\def\cdmp{cos\Delta m(t_1+t_2)}
\def\cdm{cos\Delta m(t_1-t_2)}

\def\g5{\gamma_5}
\def\gm{(1-\gamma_5)}
\def\gp{(1+\gamma_5)}

\def\mvec#1{\vec{#1}\,}
\def\pslash{\mbox{/\llap p}}
\def\slash#1{\mbox{/\llap #1}}

\def\msmall#1{\mbox{\rm \small #1}}
\newcommand{\matel}[3]{\langle #1|#2|#3\rangle}
\newcommand{\hscale}{\mu\ind{hadr}}
\newcommand{\aver}[1]{\langle #1\rangle} 
\renewcommand{\Im}{\mbox{Im}\,}
\renewcommand{\Re}{\mbox{Re}\,}
\newcommand{\GeV}{\,\mbox{GeV}}
\newcommand{\MeV}{\,\mbox{MeV}}
\newcommand{\BR}{\,\mbox{BR}}
\newcommand{\dd}{{\rm d}}
\def\d{{\bf d}}
\def\V{{\bf V}}